\documentclass[vecphys]{svmult}

\usepackage{makeidx}         
\usepackage{graphicx}        
\usepackage{multicol}        
\usepackage{amssymb}

\makeindex             

\begin{document}

\title*{Polarons: from single polaron to short scale phase separation.}
\author{V.V. Kabanov\inst{1}}
\institute{J. Stefan Institute, Jamova 39, 1001, Ljubljana,
Slovenia \texttt{viktor.kabanov@ijs.si}}
%
%
\maketitle

\section{Introduction}
\label{sec:1} There is substantial evidence that the ground state
in many of oxides is inhomogeneous\cite{egami}. In cuprates, for
example neutron-scattering experiments suggest that phase
segregation takes place in the form of stripes or short segments
of stripes\cite{bianconi,mook}. There is some controversy whether
this phase segregation is associated with magnetic interactions.
On the other hand it is also generally accepted that the charge
density in cuprates is not homogeneous.

The idea of charge segregation has a quite long history (see for
example\cite{zaanen,emery,gorkov}).  In charged systems the phase
separation is often accompanied by the charge segregation.
Breaking of the charge neutrality leads to the appearance of the
electric field and substantial contribution of the electrostatic
energy to the thermodynamic potential\cite{gorkov2,low}. Recently
it was suggested that interplay of the short range lattice
attraction and the long-range Coulomb repulsion between charge
carriers could lead to the formation  of short metallic \cite
{fedia1,akpz} or insulating\cite{akpz,fedia2} stripes of polarons.
If the attractive potential is isotropic, charged bubbles have a
spherical shape. Kugel and Khomskii \cite{klim} suggested recently
that the anisotropic attraction forces caused by Jahn-Teller
centers could lead to the phase segregation in the form of
stripes. The long-range anisotropic attraction forces appear as
the solution of the full set of elasticity equations (see
ref.\cite{eremin}). Alternative approach to take into account
elasticity potentials was proposed in ref.\cite{Shenoy} and is
based on the proper consideration of compatibility constraint
caused by absence of a dislocation in the solid. Phenomenological
aspects of the phase separation was discussed recently in the
model of Coulomb frustrated phase transition \cite{lorenzana,
spivak, mertelj}.

Here we consider some aspects of the phase separation associated
with different types of ordering of $charged$ polarons. The
formation of polaronic droplets in this case is due to competition
of two types of interactions: the long range Coulomb repulsion and
attraction generated by the deformation field. It is important to
underline that if we consider the system of neutral polarons
(without Coulomb repulsion), it shows a first order phase
transition at constant chemical potential, and is unstable with
respect to global phase separation at fixed density\cite{mertelj}.
Electron-phonon interaction may be short range and long range,
depending on the type of phonons involved. In most of the cases we
consider phonons of the molecular type, leading to short range
forces. In some cases we consider long-range Fr\"ohlich
electron-phonon interaction and interaction with the strain.

 \section{Single polaron in the adiabatic
approximation}.

\bigskip

Adiabatic theory of polarons was formulated many years ago
\cite{holstein,rashba} and we briefly formulate here the main
principles of adiabatic theory for the particular case of
interaction with molecular vibrations (Holstein model). The
central equation in the adiabatic theory is the Schr\"odinger
equation for the electron in the external potential of the
deformation field. In the discrete version it has the following
form \cite{kabmash}:
\begin{equation}
-\sum_{{\bf m}\neq 0}t({\bf m})[ \psi_{\bf n}^{k}-\psi_{\bf
n+m}^{k}]+\sqrt{2} g \omega_{0} \varphi_{\bf n}
 \psi_{\bf n}^{k} = E_{k}\psi_{\bf n}^{k}.
\end{equation}
Here $t({\bf m})$ is hopping integral, $\psi_{\bf n}$ is
electronic wave function on the site ${\bf n}$, $\varphi_{\bf n}$
is the deformation at the site ${\bf n}$, $g$ is electron phonon
coupling constant and $\omega_{0}$ is phonon frequency, $k$
describes quantum numbers of the problem. Important assumption of
the adiabatic approximation is that deformation field is very slow
and we assume that $\varphi_{\bf n}$ is time independent $\partial
\varphi/\partial t =0$  when we substitute it to the Schr\"odinger
equation for electron. Therefore, $\omega_{0}\rightarrow 0$ and
$g^{2}\rightarrow\infty$ but the product $g^{2}\omega_{0}=E_{p}$
is finite and called polaron shift. The equation for $\varphi_{\bf
n}$ has the form \cite{kabmash}:
\begin{equation}
\varphi_{\bf n}=-\sqrt{2} g|\psi_{\bf n}^{0}|^{2}
\end{equation}
Here $\psi_{\bf n}^{0}$ corresponds to the ground state solution
of the Eq.(1). After substitution of the Eq.(2) to Eq.(1) we
obtain:
\begin{equation}
-\sum_{{\bf m}\neq 0}t({\bf m})[ \psi_{\bf n}^{k}-\psi_{\bf
n+m}^{k}]-2 E_{p} |\psi_{\bf n}^{0}|^{2}
 \psi_{\bf n}^{k} = E_{k}\psi_{\bf n}^{k}.
\end{equation}
As a result nonlinear Schr\"odinger equation (Eq.(3)) describes
the ground state of the polaron. All excited states are the
eigenvalues and eigenfunctions of the $linear$ Schr\"odinger
equation in the presence of the external field determined by the
deformation field Eq.(2). Polaron energy is the sum of two
contributions. The first is the energy of electron in the
self-consistent potential well, determined by Eq.(3), and the
second is the energy of the strain field $\varphi$ itself
$E_{pol}=E_{0}+\omega_{0}\sum_{\bf n}\varphi_{\bf n}^{2}/2$. The
polaron energy is presented in the Fig.1 as a function of the
dimensionless coupling constant $2g^{2}\omega_{0}/t$ in 1D, 2D and
3D cases.
\begin{figure}
\centering
\includegraphics[height=6cm]{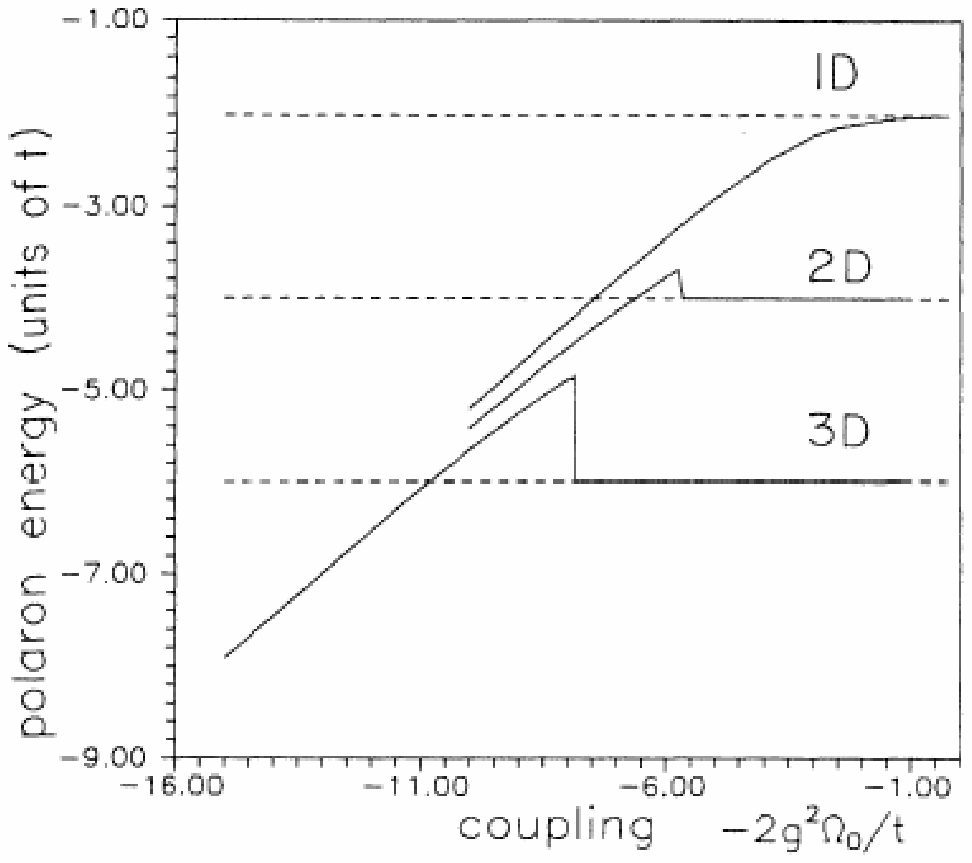}
\caption{Polaron energy as a function of coupling constant in
1D,2D and 3D cases. Dashed lines represent the energy of
delocalized solution in 1D, 2D and 3D respectively}
\label{fig:1}
\end{figure}

There is very important difference between 1D and 2,3D cases.
Polaron energy for 1D case is always less then the energy of the
delocalized state(dashed line). Polaron is always stable in 1D. In
2D and 3D cases there is critical value of the coupling constant
where first localized solution of Eq.(3) appears. It is
interesting that the energy of the solution is higher then energy
of the delocalized state. Therefore in 2D and 3D cases there is
range of coupling constant where polaron is metastable.
Delocalized solution is always stable in 3D case. Therefore, the
barrier, which separates localized and delocalized states exists
in the whole region of the coupling constants, where self-trapped
solution exists. In 2D case the delocalized state is unstable at
large value of the coupling constant. The barrier separating
localized and delocalized states forms only in the restricted
region of the coupling constant \cite{kabmash}. To demonstrate
that we have plotted polaron energy as a function of its radius in
2D in the Fig.2. As it is clearly seen from this figure, barrier
has disappeared at $g > g_{c3}=\sqrt{2\pi t/\omega_{0}}$
\cite{kabmash}.
\begin{figure}
\centering
\includegraphics[height=6cm]{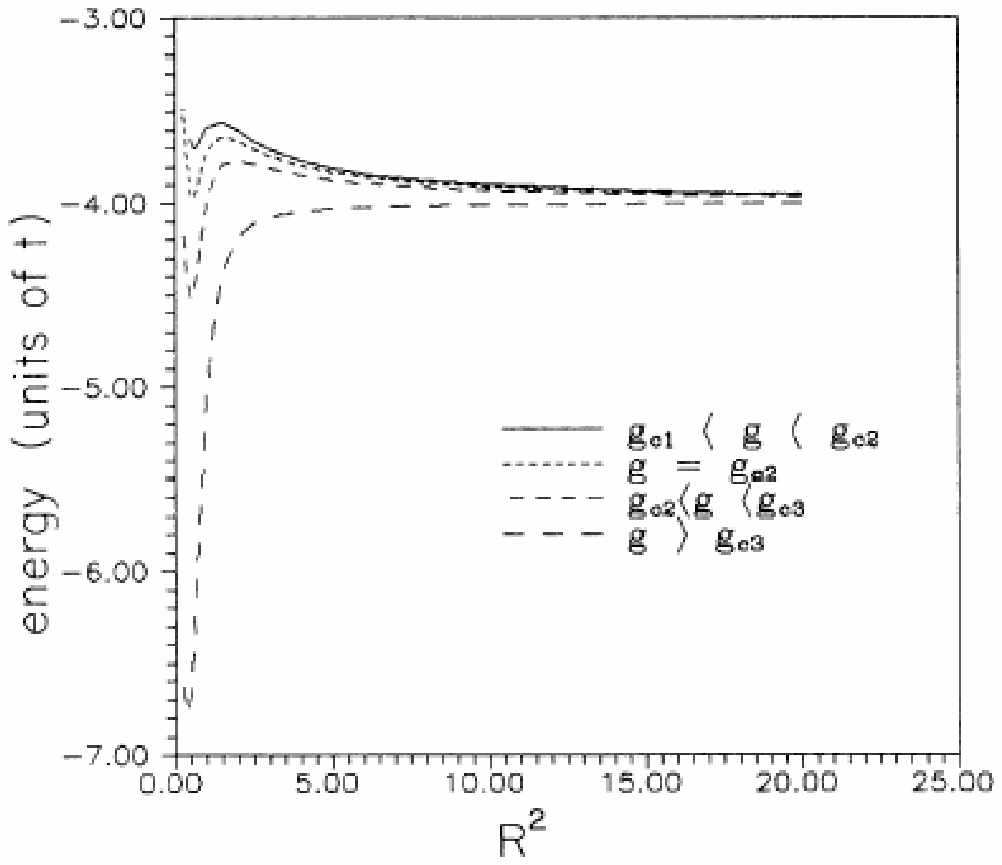}
\caption{Polaron energy as a function of radius in 2D. For
$g>g_{c3}$ $\partial E_{pol}/\partial R <0$.
$g_{c1}=1.69\sqrt{t/\omega_{0}}$, and
$g_{c2}=1.87\sqrt{t/\omega_{0}}$} \label{fig:2}
\end{figure}

There are two types of non-adiabatic corrections to adiabatic
polaron. The first is related to renormalization of local phonon
modes. Fast motion of the electron within polaronic potential well
leads to the shift of the local vibrational frequency:
\begin{equation}
\omega=\omega_{0}[1-zt^{2}/2(g^2\omega_{0})^{2}]^{1/2}
\end{equation}
here $z$ is the number of nearest neighbors. This formula is valid
in the strong coupling limit $g^2\omega_{0}\gg t$ and when
tunneling frequency of the polaron is much smaller then phonon
frequency.

Another type of correction is related to the restoration of
translational symmetry (Goldstone mode) and describes polaron
tunnelling and formation of the polaron band. This correction was
calculated in the original paper of Holstein
(Ref.\cite{holstein}). Slightly improved formula was derived in
\cite{alkabray}. In the adiabatic limit polaron tunnelling is
exponentially suppressed $t_{eff} \propto
\sqrt{E_{p}\omega_{0}}\exp{(-g^{2})}$ (see Eq. (9) of the
Ref.\cite{alkabray}) and should be smaller then phonon frequency
$\omega_{0}$.

In the following we will neglect all nonadiabatic corrections. We
will consider polaron as pure localized state, and all corrections
which contain phonon frequency itself and tunnelling amplitude for
polaron are neglected.

\section{Strings  in  charge-transfer Mott insulators: effects
of lattice vibrations and the Coulomb interaction}


Here we prove that the Fr\"ohlich electron-phonon interaction
combined with the direct Coulomb repulsion  does not lead to
charge  segregation like  strings in doped narrow-band insulators,
both in the nonadiabatic and adiabatic regimes.
 However, this interaction significantly reduces  the Coulomb repulsion, which
might allow much weaker antiferromagnetic and/or short-range
electron-phonon interactions to  segregate charges in the doped
insulators, as suggested by previous studies
\cite{zaanen,emery,fedia2}.

To begin with, we consider a generic Hamiltonian, including,
respectively,  the kinetic energy of carriers, the Fr\"ohlich
electron-phonon interaction, phonon energy, and the  Coulomb
repulsion as
\begin{eqnarray}
 H &=& \sum_{i \neq j}t({\bf m-n})\delta_{s,s'}
 c^{\dagger}_{i}c_{j}+\sum_{{\bf q},
 i}\omega_{\bf q}n_{i}\left[u_{i}({\bf q})d_{\bf q} + H.c.\right] \cr
 &+&
 \sum_{\bf q}\omega_{\bf q}(d_{\bf q}^{\dagger}d_{\bf q}+1/2)+
 {1\over{2}}\sum_{i\neq j}V({\bf m-n})n_{i}n_{j}
 \end{eqnarray}
with  bare hopping integral $t({\bf m })$, and
 matrix element of the electron-phonon interaction
 \begin{equation}
 u_{i}({\bf q})={1\over{\sqrt{2N}}}\gamma({\bf q})e^{i{\bf q\cdot
 m}}.
 \end{equation}
 Here $i=({\bf m},s)$, $j=({\bf n},s')$ include  site ${\bf m,n}$
 and spin ${s,s'}$ quantum numbers,
$n_{i}=c^{\dagger}_{i}c_{i}$,  $c_{i},d_{\bf q}$ are the electron
(hole) and phonon operators, respectively, and $N$ is the number
of sites.  At large distances ( or small $q$)
 one finds
\begin{equation}
\gamma({\bf q})^2\omega_{\bf q}={4\pi  e^2\over{\kappa q^2}},
\end{equation}
 and
\begin{equation}
V({\bf m-n})={e^2\over{\epsilon_{\infty} |{\bf m-n}|}}.
\end{equation}
The phonon frequency $\omega_{\bf q}$ and the static and
high-frequency dielectric constants in $\kappa^{-1}=
\epsilon_{\infty}^{-1}-\epsilon_{0}^{-1}$ are those of the host
insulator ($\hbar=c=1$).

 In the adiabatic limit one can apply a discrete
version  of the continuos nonlinear  equation \cite{pek} proposed
for the Holstein model (Eq.(1)), extended to the case of the
deformation and Fr\"ohlich interactions in Ref.
\cite{fedia2,fedia1,akpz}. Applying the Hartree approximation for
the Coulomb repulsion,  the single-particle wave-function,
$\psi_{\bf n}$ (the amplitude of the Wannier state $|{\bf
n}\rangle$) obeys the following equation
\begin{equation}
-\sum_{{\bf m}\neq 0}t({\bf m})[ \psi_{\bf n}-\psi_{\bf
n+m}]-e\phi_{\bf n}
 \psi_{\bf n} = E\psi_{\bf n}.
\end{equation}

The potential $\phi_{{\bf n},k}$ acting on a fermion $k$ at the
site ${\bf n}$  is  created by the polarization of the lattice
$\phi_{{\bf n},k}^{l}$ and by the Coulomb repulsion with the other
$M-1$ fermions, $\phi_{{\bf n},k}^{c}$,
\begin{equation}
\phi_{{\bf n},k} = \phi_{{\bf n},k}^{l}+\phi_{{\bf n},k}^{c}.
\end{equation}
Both potentials satisfy the descrete Poisson equation as
\begin{equation}
\kappa \Delta \phi_{{\bf n},k}^{l} = 4\pi e
\sum_{p=1}^{M}|\psi_{{\bf n},p}|^{2},
\end{equation}
and
\begin{equation}
\epsilon_{\infty}\Delta \phi_{{\bf n},k}^{c} = -4\pi e \sum_{p=1,p
\neq k}^{M}|\psi_{{\bf n},p}|^{2},
\end{equation}
with $\Delta \phi_{\bf n}=\sum_{\bf m}(\phi_{\bf n}-\phi_{\bf
n+m})$. Differently from Ref. \cite{fedia1} we include the Coulomb
interaction in Pekar's functional $J$ \cite{pek}, describing the
total energy, in a selfconsistent manner using the Hartree
approximation, so that\cite{akpz}
\begin{eqnarray}
J&=&-\sum _{{\bf n},p,{\bf m}\neq 0}\psi^*_{{\bf n},p}t({\bf m}) [
\psi_{{\bf n},p}-\psi_{{\bf n+m},p}]\cr &-&{2\pi
e^{2}\over{\kappa}}\sum_{{\bf n},p,{\bf m},q}|\psi_{{\bf
n},p}|^{2}\Delta^{-1} |\psi_{{\bf m},q}|^{2}\cr & +& {2\pi
e^{2}\over{\epsilon_{\infty}}}\sum_{{\bf n},p,{\bf m},q \neq
p}|\psi_{{\bf n},p}|^{2}\Delta^{-1} |\psi_{{\bf m},q}|^{2}.
\end{eqnarray}

If we assume, following Ref. \cite{fedia2} that the
single-particle function of a fermion trapped in a string  of the
length $N$ is a simple exponent, $\psi_{n}=N^{-1/2} \exp (ikn)$
with the periodic boundary conditions, then the functional $J$ is
expressed as $J=T+U$, where $T=-2t(N-1)\sin(\pi
M/N)/[N\sin(\pi/N)]$ is the kinetic energy (for an $odd$ number
$M$ of spinless fermions),  proportional to $t$, and
\begin{equation}
U=-{e^{2}\over{\kappa}} M^{2} I_{N} +
{e^{2}\over{\epsilon_{\infty}}} M(M-1) I_{N},
\end{equation}
corresponds to the polarisation and the Coulomb energies.  Here
the integral $I_{N}$ is given by
\begin{eqnarray}
 I_N&=& {\pi \over{(2\pi)^3}} \int _{-\pi}^\pi dx \int _{-\pi}^\pi dy  \int
     _{-\pi}^\pi dz {\sin(Nx/2)^2\over{N^2 \sin(x/2)^2}}\cr
&\times&( 3-\cos x -\cos
     y -\cos z)^{-1}.
\end{eqnarray}
$I_{N}$ has the following asymptotic \cite{akpz}:
\begin{equation}
I_{N}= {1.31+\ln N\over{N}},
\end{equation}
 The asymptotic is derived also analyticially at large $N$ by the use of the
 fact that
 $ \sin(Nx/2)^2/(2\pi N\sin(x/2)^2)$ can be replaced by a $\delta$- function.
 If we split the first (attractive) term in Eq.(14) into two parts by
replacing $M^2$ for $M+M(M-1)$, it  becomes clear that the net
interaction between polarons remains repulsive
 in the adiabatic regime because $\kappa
 >\epsilon_{\infty}$. Hence, there are no strings
 within the Hartree approximation for the Coulomb
 interaction. Strong
 correlations  do not change this conclusion. Indeed,
if we take  the Coulomb energy of spinless one-dimensional
fermions comprising both
 Hartree and exchange terms as
\begin{equation}
  E_C={e^2M(M-1)\over{N\epsilon_{\infty}}} [0.916+\ln M],
\end{equation}
the polarisation and Coulomb  energy  per particle  becomes (for
large $M>>1$)
\begin{equation}
U/M={e^2M\over{N \epsilon_{\infty}}}[0.916+\ln M - \alpha(1.31+\ln
N)],
\end{equation}
where $\alpha= 1-\epsilon_{\infty}/\epsilon_0 <1$. Minimising this
energy with respect to the length of the string $N$ we find
\begin{equation}
N= M^{1/\alpha} \exp (-0.31+0.916/ \alpha),
\end{equation}
and
\begin{equation}
(U/M)_{min}= -{e^2\over{\kappa}} M^{1-1/\alpha} \exp(0.31-0.916
/\alpha).
\end{equation}
Hence,  the potential energy per particle increases
 with the number of particles so that the
energy of $M$ well separated polarons is lower than the energy of
polarons trapped in a string no matter whether they are correlated
or not. The opposite conclusion of Ref. \cite{fedia1} originates
in an incorrect approximation of the integral
 $I_N \propto N^{0.15}/N$. The correct asymptotic result is
 $I_N = \ln(N)/N$.

 One can argue \cite{fedia1} that a  finite  kinetic
energy ($t$) can stabilise a string of a finite length.
Unfortunately, this is not correct either. We performed exact
(numerical) calculations of the total energy $E(M,N)$ of $M$
spinless fermions in a string of the length $N$ including both
kinetic  and potential energy  with the typical values of
$\epsilon_{\infty}=5$ and $\epsilon_{0}=30$. The  local energy
minima (per particle) in the string of the length $1 \leq N \leq
69$ containing $M \leq N/2$ particles are presented in the Table
1. Strings with even fermion numbers carry a finite current and
hence the local minima are found for odd $M$.
   In the  extreme wide
band regime with $t$ as large as 1 eV  the global string  energy
minimum is found at $M=3, N=25$ ($E= -2.1167$ eV),  and at $M=3,
N=13$ for $t=0.5$ eV ($E= -1.2138$ eV). However, this is $not$ the
ground state energy in both cases. The energy of well separated
$d\geq 2$-dimensional polarons is well below, less than $-2dt$ per
particle (i.e.  $-6$ eV in the first case and $-3$ eV in the
second one in the three dimensional cubic lattice, and $-4$ eV and
$-2$ eV, respectively, in the two-dimensional square lattice).
This  argument is applied for any values of $\epsilon_{0},
\epsilon_{\infty}$ and $t$.  As a result we have proven that
strings are impossible with the Fr\"ohlich interaction alone
contrary to the erroneous Ref. \cite{fedia1}.
\begin{table}
\centering \caption{$E(M,N)$ for $t=1$ eV and $t=0.5$ eV}
\label{tab:1}
\begin{tabular}{lllll}
\hline\noalign{\smallskip}
 & $t=1$eV &  & $t=0.5$eV & \\ \hline
M \hspace{1cm}& N  & E(M,N) \hspace{0.5cm} &   N   & E(M,N)\\
\noalign{\smallskip}\hline\noalign{\smallskip}
1 \hspace{1cm}& 11 & -2.0328 \hspace{0.5cm}&   3   & -1.1919 \\
3 \hspace{1cm}& 25 & -2.1167 \hspace{0.5cm}&   13  & -1.2138 \\
5 \hspace{1cm}& 42 & -2.1166 \hspace{0.5cm}&   25  & -1.1840 \\
7 \hspace{1cm}& 61 & -2.1127 \hspace{0.5cm}&   40  & -1.1661 \\
\noalign{\smallskip}\hline
\end{tabular}
\end{table}

The Fr\"ohlich interaction is, of course, not the only
electron-phonon interaction in  ionic solids. As discussed in Ref.
\cite{alemot}, any short range electron-phonon interaction, like,
for example, the  Jahn-Teller (JT) distortion  can overcome
 the residual weak repulsion
of Fr\"ohlich  polarons to form small bipolarons. At large
distances small nonadibatic bipolarons weakly  repel each other
due to the long-range Coulomb interaction, four times of that of
polarons, Eq.(9).   Hence, they  form a liquid state
\cite{alemot}, or bipolaronic-polaronic crystal-like structures
\cite{aub} depending on their effective mass and density. The
fact, that the Fr\"ohlich interaction almost nullifies the Coulomb
repulsion in oxides justifies the use of the Holstein-Hubbard
model \cite{bis,feh}. The ground state of the 1D Holstein-Hubbard
model is a liquid of intersite bipolarons with a significantly
reduced mass (compared with the on-site bipolaron) as  shown
recently \cite{bon}. The bound states of three or more polarons
are not stable in this model, thus ruling out phase separation.
However, the situation might be different if the antiferromagnetic
\cite{zaanen,emery} and JT interaction\cite{gorkov} or any other
short (but finite) range electron-phonon interaction   are strong
enough.  Due to long-range nature of the Coulomb repulsion the
length of a string should be finite (see, also
Ref.\cite{fedia2,akpz,bia}).

To summarize we conclude that there are no strings in ionic doped
insulators with the  Fr\"ohlich interaction alone. Depending on
their density and mass polarons remain in a liquid state or Wigner
crystal. On the other hand the short-range electron-phonon  and/or
antiferromagnetic interactions might  provide a liquid bipolaronic
state and/or   charge segregation (strings of a finite length)
since the long-range Fr\"ohlich interaction significantly reduces
the Coulomb repulsion in highly polarizable ionic insulators.


\section{Ordering of charged polarons: Lattice gas model}


In this section we consider macroscopic system of polarons in the
thermodynamic limit. To underline nontrivial geometry of the phase
separation we consider two-fold degenerate electronic states which
interact with nonsymmetric deformation field. In our derivation we
follow particular model for high-T$_{c}$ superconductors.
Nevertheless the results are general enough and are applicable to
many Jahn-Teller systems.

Recently we formulated the model\cite{mk} where we suggested that
interaction of a two-fold degenerate electronic state with fully
symmetric of the small group $\tau_{1}$ phonon modes at a finite
wave-vector can lead to a local nonsymmetric deformation and
short-length scale charge segregation in high-T$_{c}$ materials.
Here we reduced the proposed model to the lattice gas
model\cite{mertelj} and showed that the model indeed displays
phase separation, which may occur in the form of stripes or
clusters depending on the anisotropy of the short range attraction
between localized carriers\cite{mertelj}. We also generalized the
model taking into account interaction of the Jahn-Teller centers
via elasticity induced field\cite{mertelj}. We showed that the
model without Coulomb repulsion displays a first order phase
transition at a constant chemical potential. When the number of
particles is fixed, the system is unstable with respect to the
global phase separation below certain critical temperature. In the
presence of the Coulomb repulsion global phase separation becomes
unfavorable due to a large contribution to the energy from long
range Coulomb interaction. The system shows mesoscopic phase
separation where the size of charged regions is determined by the
competition between the energy gain due to ordering and energy
cost due to breaking of the local charge neutrality. Since the
short range attraction is anisotropic the phase separation may be
in the form of short segments or/and stripes.

Let us start with the construction of a real-space Hamiltonian
which couples 2-fold degenerate electronic states (or
near-degenerate states) with optical phonons of $\tau_{1}$
symmetry. Two-fold degeneracy is essential because in that case
formation of the polaronic complexes leads to reduction not only
of translational symmetry but also reduction of the point group
symmetry. Since the Hamiltonian needs to describe a 2-fold
degenerate system, the 2-fold degenerate states - for example the
two $E_{u}$ states corresponding to the planar hybridized Cu
$d_{x^{2}-y^{2}}$ , O $p_{x}$ and $p_{y}$ orbitals, or the $E_{u}$
and $E_{g}$ states of the apical O - are written in the form of
Pauli matrices $\sigma _{i}$. Taking into account that the states
are real, the Pauli matrices which describe transitions between
the levels transform as $A_{1g}$ ($k_{x}^{2}+k_{y}^{2}$) for
$\sigma _{0}$, $B_{1g}$ ($k_{x}^{2}-k_{y}^{2}$) for $\sigma _{3}$,
$B_{2g}$ ($k_{x}k_{y}$) for $\sigma _{1}$, and $A_{2g}$ ($s_{z}$)
\ for $\sigma _{2}$ representations respectively. Collecting terms
together by symmetry we can construct \emph{effective}
electron-spin-lattice interaction Hamiltonian given in the
Ref.\cite{mk}.  Here we consider a simplified version of the JT
model Hamiltonian \cite{mk}, taking only the deformation of the
$B_{1g}$ symmetry:
\begin{equation}
H_{JT}=g\sum_{\mathbf{n},\mathbf{l}}\sigma _{3,\mathbf{l} }f(\mathbf{n}%
)(b_{\mathbf{l+n}}^{\dagger }+ b_{\mathbf{l+n}}),
\end{equation}
here the Pauli matrix $\sigma _{3,\mathbf{l}}$ describes two
components of the electronic doublet, and
$f(\mathbf{n})=(n_{x}^{2}-n_{y}^{2})f_{0}(n)$ where $f_{0}(n)$ is
a symmetric function describing the range of the interaction. For
simplicity we omit the spin index in the sum. The model could be
easily reduced to a lattice gas model\cite{mertelj}. Let us
introduce the classical variable $\Phi _{\mathbf{i}
}=<b_{\mathbf{i}}^{+}+b_{\mathbf{i}}>/\sqrt{2}$ and minimize the
energy as a function of $\Phi _{\mathbf{i}}$ in the presence of
the harmonic term $\omega \sum_{\mathbf{i}}\Phi
_{\mathbf{i}}^{2}/2$. We obtain the deformation, corresponding to
the minimum of energy,
\begin{equation}
\Phi_{\mathbf{i}}^{(0)}=-\sqrt{2}g/\omega \sum_{\mathbf{n}}\sigma
_{3, \mathbf{i}+\mathbf{n}}f(\mathbf{n}).
\end{equation}
Substituting $\Phi_{\mathbf{i}}^{(0)}$ into the Hamiltonian (1)
and taking into account that the carriers are charged we arrive at
the lattice gas model. We use a pseudospin operator $S=1$ to
describe the occupancies of the two electronic levels $n_{1}$and
$n_{2}$. Here $S^{z}=1$
corresponds to the state with $n_{1}=1$ , $n_{2}=0$, $S_{i}^{z}=-1$ to $%
n_{1}=0$, $n_{2}=1$ and $S_{i}^{z}=0$ to $n_{1}=n_{2}=0$.
Simultaneous occupancy of both levels is excluded due to a high
on-site Coulomb repulsion energy. The Hamiltonian in terms of the
pseudospin operator is given by\cite{mertelj}
\begin{equation}
H_{LG}=\sum_{\mathbf{i},\mathbf{j}}(-V_{l}(\mathbf{i}-\mathbf{j})S_{\mathbf{%
i }}^{z}S_{\mathbf{j}}^{z}+V_{c}(\mathbf{i}-\mathbf{j})Q_{\mathbf{i}}Q_{%
\mathbf{j}}),
\end{equation}
where $Q_{\mathbf{i}}=(S_{\mathbf{i}}^{z})^{2}$. $V_{c}(\mathbf{n}%
)=e^{2}/\epsilon_{0}a(n_{x}^{2}+n_{y}^{2})^{1/2}$ is the Coulomb
potential, $e$ is the charge of electron, $\epsilon _{0}$ is the
static dielectric constant and $a$ is the effective unit cell
period. The anisotropic short range attraction potential is given
by $V_{l}(\mathbf{n})=g^{2}/\omega \sum_{\mathbf{m}
}f(\mathbf{m})f( \mathbf{n}+\mathbf{m})$. The attraction in this
model is generated by the interaction of electrons with optical
phonons. The radius of the attraction force is determined by the
radius of the electron-phonon interaction and the dispersion of
optical phonons\cite{akpz}.

A similar model can be formulated in the limit of continuous
media\cite{mertelj}. The deformation is characterized by the
components of the strain tensor. For the two dimensional case we
can define 3 components of the strain tensor:
$e_{1}=u_{xx}+u_{yy}$, $\epsilon =u_{xx}-u_{yy}$ and
$e_{2}=u_{xy}$ transforming as the $A_{1g}$, $B_{1g}$ and $B_{2g}$
representations of the $D_{4h}$ group respectively. These
components of the tensor are coupled linearly with the two-fold
degenerate electronic state which transforms as the $E_{g}$ or
$E_{u}$ representation of the point group. Similarly to the case
of previously considered interaction with optical phonons we keep
the interaction with the deformation $\epsilon $ of the $B_{1g}$
symmetry only. The Hamiltonian without the Coulomb repulsion term
has the form:
\begin{equation}
H=g\sum_{\mathbf{i}}S_{\mathbf{i}}^{z}\epsilon
_{\mathbf{i}}+\frac{1}{2} \left(
A_{1}e_{1,\mathbf{i}}^{2}+A_{2}\epsilon
_{\mathbf{i}}^{2}+A_{3}e_{3, \mathbf{i}}^{2}\right),
\end{equation}
where $A_{j}$ are corresponding components of the elastic modulus
tensor. The components of the strain tensor are not independent
\cite{Bishop,Shenoy} and satisfy the compatibility condition:
\[
\nabla^{2} e_{1}({\bf r})-4\partial^{2} e_{2}({\bf r})/\partial x
\partial y = (\partial^{2}/\partial x^{2} - \partial^{2}/\partial
y^{2}) \epsilon({\bf r})
\]
The compatibility condition leads to a long range anisotropic
interaction between polarons. The Hamiltonian in the reciprocal
space has the form:
\begin{equation}
H=g\sum_{\mathbf{k}}S_{\mathbf{k}}^{z}\epsilon _{\mathbf{k}}+(A_{2}+A_{1}U(%
\mathbf{k}))\frac{\epsilon _{\mathbf{k}}^{2}}{2}.
\end{equation}
The Fourrier transform of the potential is given by:
\begin{equation}
U(\mathbf{k})=\frac{(k_{x}^{2}-k_{y}^{2})^{2}}{
k^{4}+8(A_{1}/A_{3})k_{x}^{2}k_{y}^{2}}.
\end{equation}
By minimizing the energy with respect to $\epsilon_{k}$ and taking
into account the long-range Coulomb repulsion we again derive
Eq.(23). The anisotropic interaction potential
$V_{l}(\mathbf{n})=-\sum_{\mathbf{k}}\exp{%
(i\mathbf{k\cdot n})}\frac{g^{2}}{2(A_{2}+A_{1}U(\mathbf{k}))}$ is
determined by the interaction with the classical deformation and
is long-range. It decays as $1/r^{2}$ at large distances in 2D.
Since at large distances the attraction forces decay faster then
the Coulomb repulsion forces the attraction can overcome the
Coulomb repulsion at short distances, leading to a mesoscopic
phase separation.

Irrespective of whether the resulting interaction between polarons
is generated by acoustic or optical phonons the main physical
picture remains the same. In both cases there is an anisotropic
attraction between polarons on short distances. The interaction
could be either ferromagnetic or antiferromagnetic in terms of the
pseudospin operators depending on mutual
orientation of the orbitals. Without loosing generality we assume that $V(%
\mathbf{n})$ is nonzero only for the nearest neighbors.

Our aim is to study the model (Eq.(23)) at constant average
density,
\begin{equation}
n=\frac{1}{N}\sum_{\mathbf{i}}Q_{\mathbf{i}},  \label{eq_npart}
\end{equation}
where $N$ is the total number of sites. However, to clarify the
physical picture it is more appropriate to perform calculations
with a fixed chemical potential first by adding the term $-\mu
\sum_{\mathbf{i}}Q_{\mathbf{i}}$ to the Hamiltonian (23).

Models such as Eq.(23) without the long-range forces, were studied
many years ago on the basis of the molecular-field approximation
in the Bragg-Williams formalism \cite{lajz,siva}. The mean-field
equations for the particle density $n$ and the pseudospin
magnetization $M=\frac{1}{N}\sum_{\mathbf{i}}S_{\mathbf{i}}^{z}$
have the form\cite{lajz}:
\begin{eqnarray}
M &=&\frac{2\sinh {(2zV_{l}M/{k}_{B}{T})}}{\exp {(-\mu /k}_{B}{T)}+2\cosh {%
(2zV_{l}M/{k}_{B}{T})}}  \label{eq_M} \\
n &=&\frac{2\cosh {(2zV_{l}M/{k}_{B}{T})}}{\exp {(-\mu /{k}_{B}T)}+2\cosh {%
(2zV_{l}M/{k}_{B}{T})}}
\end{eqnarray}
here $z=4$ is the number of the nearest neighbours for a square
lattice in 2D and ${{k}_{B}}${\ is the Boltzman constant}. A phase
transition to an ordered state with finite $M$ may be of either
first or second order, depending on the value of $\mu $. For the
physically important case $-2zV_{l}<\mu <0,$ ordering occurs as a
result of the first order phase transition. The two solutions of
Eqs.(28,29) with $M=0$ and $M$ $\neq 0$ correspond to two
different minima of the free energy. The temperature of the phase
transition $T_{crit}$ is determined by the condition: $F(M=0,\mu
,T)=F(M,\mu ,T)$ where $M$ is the solution of Eq. (28). When the
number of particles is fixed (Eq.(29)), the system is unstable
with respect to global phase separation below $T_{crit}$. As a
result, at fixed $n$ two phases coexist with $n_{0}=n(M=0,\mu,T)$
and $n_{M}=n(M,\mu ,T)$, resulting in a liquid-gas-like phase
diagram (Fig.3).

To investigate the effects of the long range-forces, we performed
Monte Carlo simulations on the system (23)\cite{mertelj}. The
simulations were performed on a square lattice with dimensions up
to $L\times L$ sites with $10\leq L\leq 100$ using a standard
Metropolis algorithm\cite{Metropolis53} in combination with
simulated annealing\cite{Kirkpatrick83}. At constant $n$ one Monte
Carlo step included a single update for each site with nonzero
$Q_{i}$, where the trial move consisted from setting $S_{z}=0\,$at
the site with nonzero $Q_{i}\,$and $S_{z}=\pm 1\,$at a randomly
selected site with zero $Q_{i}$. A typical simulated annealing run
consisted from a sequence of Monte Carlo simulations at different
temperatures. At each temperature the equilibration phase
($10^{3}-10^{6}$ Monte Carlo steps) was followed by the averaging
phase with the same or greater number of Monte Carlo steps.
Observables were measured after each Monte Carlo step during the
averaging phase only. For $L \gtrsim 20$ we observe virtually no
dependence of the results on the system size.

Comparing the Monte Carlo\ results in the absence of Coulomb
repulsion shown by $t_{crit}~$in Fig. 3 with MF\ theory we find
the usual reduction of $t_{crit}$ due to fluctuations in 2D by a
factor of $\sim 2$.

\begin{figure}
\centering
\includegraphics[height=6cm]{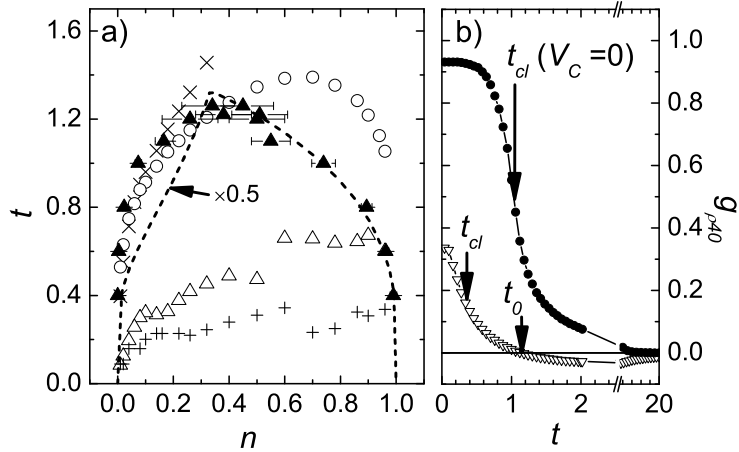}
\caption{a) The phase diagram generated by $H_{JT}$ (23) with, and
without the Coulomb repulsion (CR). The dashed line is the MF
critical temperature,  while the full triangles ($\blacktriangle
$) represent the Monte Carlo critical temperature, $t_{crit}$,
\emph{without CR}. The open circles ($\circ$) represent $t_{cl}$,
\emph{without CR}. The open triangles ($\triangle $)\ represent
$t_{cl}$ while the diagonal crosses ($\times $) represent the
onset of clustering, $t_{0}$, \emph{in the presence of }CR. The
cluster-ordering temperature (see text), $t_{co}$, (also incl. CR)
is shown as crosses (+). The size of the symbols corresponds to
the error bars. b) Typical temperature dependencies of the nearest
neighbor density correlation function $g_{\rho L}\,$for $n=0.18$
\emph{in the absence of }CR ($\bullet $) and \emph{in the presence
of }CR ($\triangledown $). Arrows indicate the characteristic
temperatures.} \label{fig:3}
\end{figure}

Next, we include the Coulomb interaction $V_{c}(r)$. We use open
boundary conditions to avoid complications due to the long range
Coulomb forces and ensure overall electroneutrality by adding a
uniformly charged background electrostatic potential (jellium) to
Eq. (23). The short range potential
$v_{l}(\mathbf{i})=V_{l}(\mathbf{i})\epsilon
_{0}a/e^{2}~$\thinspace was taken to be nonzero only for
$\left\vert \mathbf{i}\right\vert <2$ and is therefore specified
only for nearest, and next-nearest neighbours as $v_{l}(1,0)$ and
$v_{l}(1,1)$ respectively.

The anisotropy of the short range potential has a profound
influence on the particle ordering. We can see this if we fix
$v_{l}(1,0)=-1$, at a density $n=0.2$ and vary the next-nearest
neighbour potential $v_{l}(1,1)$ in the range from $-1$ to $1.$
When $v_{l}(1,1)<0$, the attraction is ''ferrodistortive''\ in all
directions, while\ for positive $v_{l}(1,1)>0$ the interaction is
''antiferrodistortive''\ along the diagonals. The resulting
clustering and ordering of clusters at $t=0.04$ is shown in Fig.
4a). As expected, a more symmetric attraction potential leads to
the formation of more symmetric clusters. On the other hand, for
$v_{l}(1,1)=1,$ the ''antiferrodistortive''\ interaction along
diagonals prevails, resulting in diagonal stripes.

In the temperature region where clusters partially order the heat
capacity ($c_L = \partial\langle E \rangle _L / \partial T$ where
$E$ is the total energy) displays  the peak at $t_{co}$. The peak
displays no scaling with $L$ indicating that no long range
ordering of clusters appears. Inspection of the particle
distribution snapshots at low temperatures (Fig. 4a) reveals that
finite size domains form. Within domains the clusters are
perfectly ordered. The domain wall dynamics seems to be much
slower than our Monte Carlo simulation timescale preventing
domains to grow. The effective $L$ is therefore limited by the
domain size. This explains the absence of the scaling and clear
evidence for a phase transition near $t_{co}$.

We now focus on the shape of the short range potential which
promotes the formation of stripes shown in Fig. 4a). We set
$v_{l}(1,0)=-1$ and $v_{l}(1,1)=0$ and study the density
dependence. Since the inclusion of the Coulomb interaction
completely suppresses the first order phase transition at
$t_{crit}$, we measure the nearest neighbor density correlation
function $g_{\rho L}=\frac{1}{4n(1-n)L^{2}}\sum_{\left|
\mathbf{m}\right| =1}\left\langle \sum_{\mathbf{i}}\left(
Q_{\mathbf{i}+\mathbf{m}}-n\right) \left( Q_{\mathbf{i}}-n\right)
\right\rangle _{L}$ to detect clustering. Here $\left\langle
{}\right\rangle _{L}$ represents the Monte Carlo average. We
define a dimensionless temperature $t_{cl}=k_{B}T_{cl}\epsilon
_{0}a/e^{2}$~as the characteristic crossover temperature related
to the formation of clusters~at which $g_{\rho L}$ rises to 50\%
of its low temperature value. The dependence of $t_{cl}$ on the
density $n$ is shown in the phase diagram in Fig. 3. Without
Coulomb repulsion $V_{c}(r)$, $t_{cl}$ follows $t_{crit}$, as
expected. The addition of Coulomb repulsion $V_{C}(r)$ results in
a significant decrease of $t_{cl}$ and suppression of clustering.
At low densities we can estimate the onset for cluster formation
by the temperature, $t_{0}$, at which $g_{\rho L}$ becomes
positive. It is interesting to note that $t_{0}$\thinspace almost
coincides with the $t_{crit}$ line at low $n$ (Fig. 3).

\begin{figure}
\centering
\includegraphics[height=6cm]{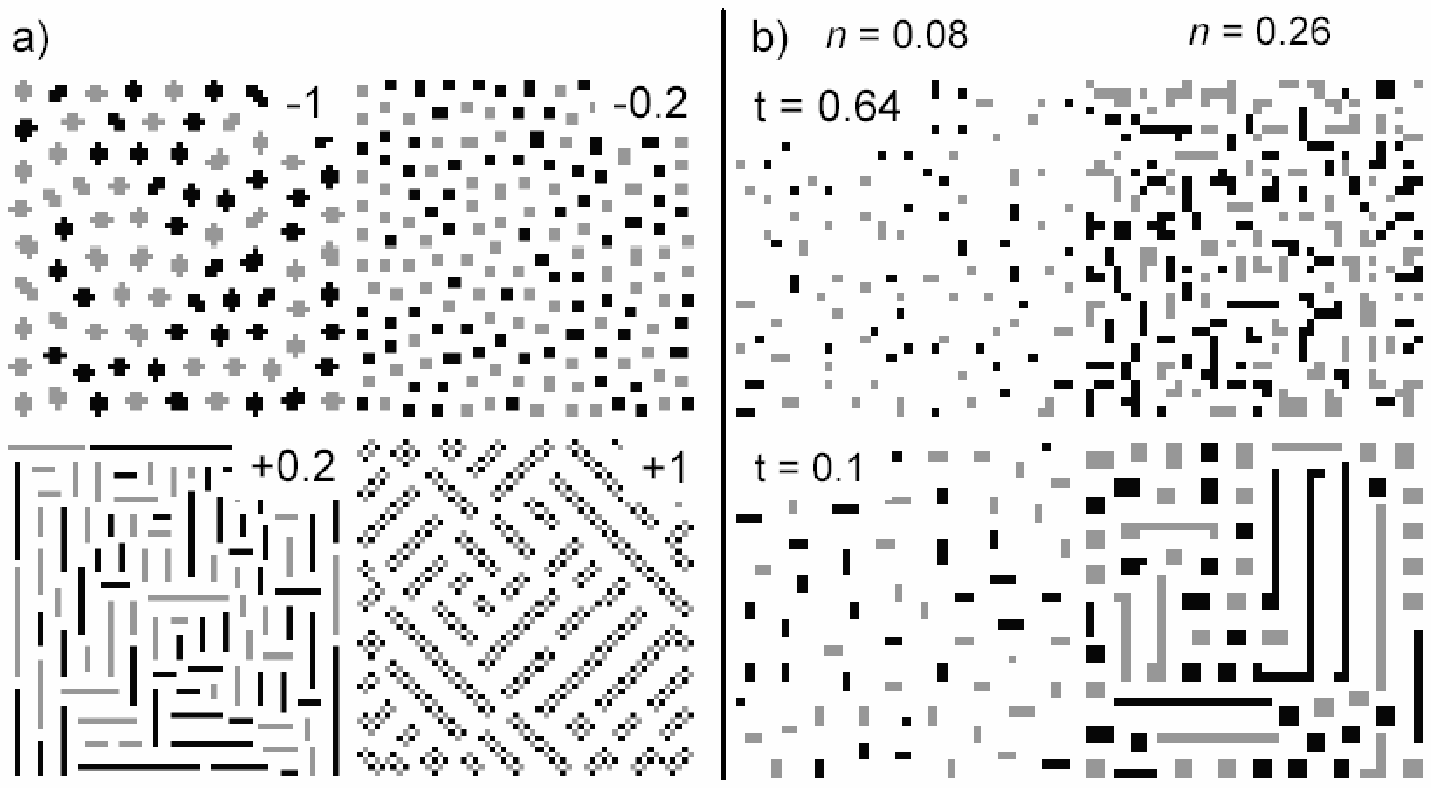}
\caption{a) Snapshots of clusters ordering at $t=0.04$, $n=0.2$
and $v_{l}(1,0)=-1$ for different diagonal $v_{l}(1,1)~$(given in
each figure). Grey and black dots represent particles clusters in
state $S_{i}^{z}=1$ and states $S_{i}^{z}=-1$ respectively. The
preference for even-particle-number clusters in certain cases is
clearly observed, for example for $v_{l}(1,1)=$ $-0.2$. b)
Snapshots of the particle distribution for two densities at two
different temperatures $t=0.64$ and $t=0.1$ respectively.}
\label{fig:4}       
\end{figure}

To illustrate this behaviour, in Fig. 4b) we show snapshots of the
calculated Monte Carlo particle distributions at two different
temperatures for different densities. The growth and ordering of
clusters with decreasing temperature is clearly observed. At low
$n$, the particles form mostly pairs with some short stripes. With
further increasing density, quadruples gradually replace pairs,
then longer stripes appear, mixed with quadruples, etc.. At the
highest density, stripes prevail forming a labyrinth-like pattern.
The density correlation function shows that the correlation length
increases with doping, but long range order is never achieved (in
contrast to the case without $V_{c}$). Note that while locally
there is no four-fold symmetry the overall correlation function
still retains 4-fold symmetry.

To get further insight in the cluster formation we measured the
cluster-size distribution\cite{mertelj}. In Fig. 5 we present the
temperature and density dependence of the cluster-size
distribution function $x_{L}(j)=\left\langle N_{p}(j)\right\rangle
_{L}/(nL^{2})$, where $N_{p}(j)$ is the total number of particles
within clusters of size $j$. At the highest temperature $x_{L}(j)$
is close to the distribution expected for the random ordering. As
the temperature is decreased, the number of larger clusters starts
to increase at the expense of single particles. Remarkably, as the
temperature is further reduced, clusters of certain size start to
prevail. This is clearly seen at higher densities (Fig.3).
Depending on the density, the prevailing clusters are pairs up to
$n \approx 0.2$, quadruples for $0.1 \lesssim n\, \lesssim 0.3$
etc.. We note that for a large range of $v_{l}(1,0),$ the system
prefers clusters with an even number of particles. Odd
particle-number clusters can also form, but have a much narrower
parameter range of stability. The preference to certain cluster
sizes becomes clearly apparent only at temperatures lower then
$t_{cl}$, and the transition is not abrupt but gradual with the
decreasing temperature. Similarly, with increasing density changes
in textures also indicate a series of crossovers.

\begin{figure}
\centering
\includegraphics[height=6cm]{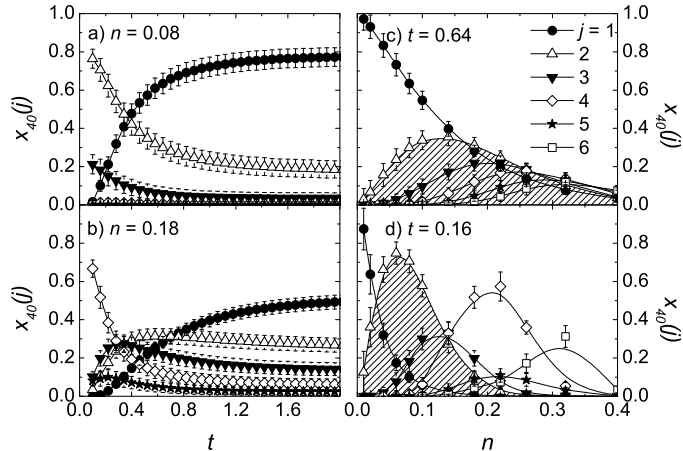}
\caption{The temperature dependence of the cluster-size
distribution function $x_{L}(j)$ (for the smallest cluster sizes)
as a function of temperature at two different average densities
$n=0.08$ (a) and $n=0.18$ (b). $x_{L}(j)$ as a function of $n$ at
the temperature between $t_{0}$ and $t_{cl}$ (c), and near
$t_{co}$ (d). The ranges of the density where pairs prevail are
very clearly seen in (d). Error bars represent the standard
deviation.} \label{fig:5}
\end{figure}

The results of the Monte Carlo simulation\cite{mertelj} presented
above allow a quite general interpretation in terms of the
kinetics of first order phase transitions\cite{landau}. Let us
assume that a single cluster of ordered phase with radius $R $
appears. As was discussed in \cite{gorkov2,mk}, the energy of the
cluster is determined by three terms: $\epsilon =-F\pi
R^{2}+\alpha \pi R+\gamma R^{3}$. The first term is the energy
gain due to the ordering phase transition where $F$ is the energy
difference between the two minima in the free energy density. The
second term is the surface energy parameterized by $\alpha $, and
the third term is the Coulomb energy, parameterized by $\gamma $.
If $\alpha <\pi F/3\gamma $, $\epsilon $ has a well defined
minimum at $R=R_{0}$ corresponding to the optimal size of clusters
in the system. Of course, these clusters are also interacting
among themselves via Coulomb and strain forces, which leads to
cluster ordering or freezing of cluster motion at low temperatures
as shown by the Monte Carlo simulations.

We conclude that a model with only anisotropic JT strain and a
long-range Coulomb interaction indeed is unstable with respect to
the short scale phase separation and gives rise to a remarkably
rich phase diagram including pairs, stripes and charge- and
orbital- ordered phases, of clear relevance to oxides. The energy
scale of the phenomena is defined by the parameters used in
$H_{JT-C}$ (23). For example, using the measured value $\epsilon
_{0}\simeq 40$~\cite{LB} for La$_{2}$CuO$_{4}$, we estimate
$V_{c}(1,0)=0.1$~eV, which is also the typical energy scale of the
''pseudogap'' in the cuprates. The robust prevalence of the paired
state in a wide region of parameters (Fig. 5 c,d) is particularly
interesting from the point of view of superconductivity. A similar
situation occurs in manganites and other oxides with the onset of
a conductive state at the threshold of percolation, but different
textures are expected to arise due to different magnitude (and
anisotropy) of $V_{l}(\mathbf{n}),$ and static dielectric constant
$\epsilon _{0}$~in the different
materials\cite{percolationmanganites}.

\section{Coulomb frustrated first order phase transition}

As it is stated in the previous section uncharged JT polarons have
the tendency to ordering. The ordering transition is a phase
transition of the first order. At the fixed density of polarons
the system is unstable with respect to the global phase
separation. The global phase separation is frustrated by charging
effects leading to short-scale phase separation. Therefore the
results of the Monte-Carlo simulation of the model (23) allow
general model independent interpretation. Let us consider the
classical free energy density corresponding to the first order
phase transition:
\begin{equation}
F_{1}=((t-1)+(\eta^{2}-1)^{2})\eta ^{2}
\end{equation}
Here $t=(T-T_{c})/(T_{0}-T_{c})$ is dimensionless temperature. At
$t=4/3$ ($T=T_{0}+(T_{0}-T_{c})/3$) the nontrivial minimum in the
free energy appears. At $t=1$ ($T=T_{0}$) the first order phase
transition occurs, but the trivial solution $\eta=0$ corresponds
to the metastable phase. At $t=0$ ($T=T_{c}$) trivial solution
becomes unstable. In order to study the case of Coulomb frustrated
phase transition we have to add coupling of the order parameter to
local charge density. Our order parameter describes sublattice
magnetization and therefore only square of the order parameter may
be coupled to the local charge density $\rho $:,
\begin{equation}
F_{coupl}=-\alpha \eta^{2}\rho
\end{equation}
The total free energy density should contain the gradient term and
the electrostatic energy:
\begin{equation}
F_{grad}+F_{el}=C(\nabla \eta )^{2}+\frac{K}{2}[\rho({\bf
r})-\bar{\rho}] \int d{\bf r}^{'}[\rho({\bf
r}^{'})-\bar{\rho}]/|{\bf r-r}^{'}|
\end{equation}
Here we write $\bar{\rho}$ explicitly to take into account global
electroneutrality. Total free energy Eqs.(30-32) should be
minimized at fixed $t$ and $\bar{\rho}$.

Let us demonstrate that Coulomb term leads to the phase separation
in the 2D case. Minimization of $F$ with respect to the charge
density $\rho({\bf r})$ leads to the following equation:
\begin{equation}
-\alpha \nabla^{2}_{3D} \eta^{2}=4\pi K [\rho({\bf
r})-\bar{\rho}]d\delta(z)
\end{equation}
here we write explicitly that density $\rho({\bf r})$ depends only
on 2D vector ${\bf r}$ and introduce layer thickness d, to
preserve correct dimensionality. Solving this equation by applying
the Furrier transform and substituting the solution back to the
free energy density we obtain:
\begin{equation}
F=F_{1}-\alpha \eta^{2}\bar{\rho}+C(\nabla \eta )^{2}-
\frac{\alpha^{2}}{8 \pi^{2} K d}\int d{\bf r}^{'}\frac{\nabla
(\eta({\bf r})^{2})\nabla (\eta({\bf r}^{'})^{2})}{|{\bf
r-r}^{'}|}
\end{equation}
As a results the free energy functional is similar to the case of
the first order phase transition with shifted critical temperature
due to the presence of the term $\alpha \eta^{2}\bar{\rho}$ and
nonlocal gradient term of higher order.

To demonstrate that uniform solution has higher energy then
nonhomogeneous solution we make Fourier transform of the gradient
term:
\begin{equation}
F_{grad} \propto Ck^{2}|\eta_{\bf k}|^{2}-\frac{\alpha^{2}k
|(\eta^{2})_{\bf k}|^{2}}{4 \pi K d}
\end{equation}
here $\eta_{\bf k}$ and $(\eta^{2})_{\bf k}$ are are Fourier
components of the order parameter and square of the order
parameter respectively. If we assume that the solution is uniform
i.e. $\eta_{0}\neq 0$ and $(\eta^{2})_{0}\neq 0$ small nonuniform
corrections to the solution reduces the free energy at small ${\bf
k}$, where second term dominates.

Proposed free energy functional is similar to that proposed in the
Ref.\cite{spivak}. The important difference is that in our case
the charge is coupled to the square of the order parameter and it
plays the role of local temperature, while in the case of
Ref.\cite{spivak} there is linear coupling of the charge to the
order parameter. Charge in that case plays the role of the
external field. Moreover, contrary to the case of Ref\cite{spivak}
where charge is accumulated near $domain$ $walls$, in our case
charge is accumulated near $interphase$ $boundaries$.

\section{Conclusion}

We have demonstrated that anisotropic interaction between
Jahn-Teller centers generated by optical and/or acoustical phonons
leads in the presence of the long range Coulomb repulsion to the
short scale phase separation. Topology of texturing differs from
charged bubbles to oriented charged stripes depending on the
anisotropy of short range potential. On the phenomenological level
inhomogeneous phase with charged regions appears due to tendency
of the system of polarons to global phase separation, while global
phase separation is frustrated by the long-range Coulomb forces.
Effectively this system may be described by standard Landau
functional with nonlocal long-range gradient term.

%
%
%
%

%
%



\printindex
\end{document}